# Evolution of Nonlinear Ion Transport in Nanopore Arrays: Ionic Conductance, Current Rectification, and Osmotic Power

**Chih-Yuan Lin**[1] **and Marija Drndić**[1*]

[1]*Department of Physics and Astronomy, University of Pennsylvania, Philadelphia, PA 19104, United States*

*email: drndic@physics.upenn.edu

## ABSTRACT

Understanding the ionic transport and scaling behaviors in nanopore arrays is essential for bridging fundamental ion physics and blue energy applications. By fabricating sub-3 nm and sub-20 nm diameter nanopore arrays (NPAs) spanning from few pores to *N* ~ 10000, we systematically investigate ionic conductance, ion current rectification, and osmotic energy conversion. We report the ionic conductance scaling laws and nonlinearity with nanopore number, with stronger deviations from linearity at lower salt concentrations. Experimental evidence reveals that surface-charge-governed conductance and ion current rectification progressively weaken with increasing *N* and even vanish as the NPA scales up to *N* ~ 10000, resulting in an underestimation of surface charge density. In a sub-3 nm densely packed array (separation ~ 25 nm), the conductance exhibits an anomalous power-law dependence on concentration, deviating markedly from the single nanopore characteristics, attributed to the strong pore interactions. Furthermore, osmotic power harvesting measurements reveal a substantial reduction in power density upon scaling, with decreases of up to three orders of magnitude over the same range. To elucidate the underlying mechanism, we developed rigorous 3D modeling showing that the nonlinear behavior originates from concentration polarization at pore entrances and suppressed electric field across NPAs, collectively hindering ion transport. Our work provides insight into nonlinear ion-transport scaling and reveals fundamental differences between transport phenomena in single nanopores and nanopore arrays.

## INTRODUCTION

Scaling of ionic conductance in a multipore system has received considerable attention over the past decade, as it plays a fundamental role in bridging microscopic physics and macroscopic membrane functions, as well as enabling the translation of laboratory-scale studies to industrial-scale applications. This connection is indispensable for the rational understanding and design of both biological systems and engineered nanodevices. The increasing interest in nanopore arrays is motivated by the growing importance of nanoporous membranes in widespread applications, including nanofiltration,[1-3] desalination,[4] blue energy conversion,[5] and biosensors[6-8].

Due to its structural flexibility and scalability, a solid-state material membrane has emerged as a robust model system to study the transport phenomena in confined spaces, explore the physics of desalination and osmotic energy, as well as mimic the functions of the biological ion channels. A few approaches have been reported to create nanometer-scaled nanopore arrays in solid-state SiN membranes, including electron beam lithography (EBL) and etching,[8-10] focus ion beam drilling,[7, 11, 12] transmission electron microscopy (TEM) drilling,[13-15] and laser drilling[16]. Among them, the EBL system is an attractive platform for large-scale fabrication due to its well-established automation, enabling the creation of large numbers of nanopores on wafer-scales. By contrast, TEMs and TEM-drilling are not yet optimized for producing large numbers of nanopores; however, TEM drilling offers precise control of sub-10 nm pore size and is capable of producing the smallest nanopores.

Solid-state nanopore arrays (NPA) have been employed as model systems to understand the ion transport behaviors as well as effects of interactions between pores at the nanoscale.[12, 15, 17] For example, in arrays with 200 nm diameter pores, the ionic conductance of NPAs was found to scale sub-linearly with increasing number of pores.[12] The degree of this sublinear behavior also depends on the geometry of the array, where the total conductance as a function of number of pores, $N$, has been shown to scale as $\sim N/\log(N)$ for a linear arrays and $\sim N^{0.5}$ for a square array. Another study using arrays of $N$ = 100 pores with 180 nm diameter experimentally showed that the pore spacing strongly influences diffusion layers and transport dynamics.[17] An overlapping diffusion profile was observed at smaller pore-pore spacings, whereas an independent diffusion profile was observed at larger pore separations. By sufficiently separating two sub-3 nm diameter pores (> 1 μm apart), the total conductance at 1M KCl was measured to be equal to the sum of individual pore conductances.[14] Later on, Cao *et al*.[15] fabricated 3 closely spaced co-linear nanopores and attributed the measured conductance suppression to the overlap of ionic depletion zones on one side of nanopores entrances (also known as access region). Authors also proposed how diode-like charge distribution on the nanopore wall is capable of tuning the effect of pore interactions on the conductance. Other studies, motivated by neuromorphic functions in nanofluidic devices such as "synaptic-like ion transport behavior" sub nanometer diameter hexagonal boron nitride (hBN) arrays were investigated theoretically.[18] The underlying mechanism was attributed to the competitive transport between potassium and sodium ions in the binary mixture of KCl and NaCl solutions. Relevant literature has been summarized in **Table 1**.

In addition to fundamental studies, nanopore arrays have been considered for applications in biomolecule sensing[7, 11, 19, 20] and osmotic energy conversion[21-24]. Previous studies revealed that the huge reduction of osmotic energy harvested in a multipore system[25] and porous membranes, such as anodic aluminum oxide[26] and Nafion[27], is restricting the availability of industrial applications. Theoretical analysis pointed out that the power density decreases significantly with increasing pore number and pore density from that obtained by a linear extrapolation of a single pore, which arises from significant concentration polarization at the membrane interface.[22, 25] The influence of pore spacing on the osmotic energy was also studied in arrays of nanopores formed using dielectric breakdown with an AFM tip to fabricate 3 by 3 arrays of sub-10 nm diameter pores in a hBN/SiN membrane.[21] Power density of 9 pores was found to decrease as pore spacing decreases, implying that pore-to-pore interaction in the packed NPA compromises the performance of osmotic power density. Despite considerable advances, the scalability and the practical viability of osmotic energy remain elusive.[28] Furthermore, understanding transport phenomena of ions and fluids in a multipore system is still in preliminary stages limited by a lack of experimental studies. A thorough investigation of ion transport behaviors within NPA is essential to bridge the gap between the single pore models and multipore systems, paving the way for the development of next-generation membranes.

Here, we perform systematic measurements of ion transport in nanopore arrays from 1 to 10000 silicon nitride nanopores of diameters from sub-3 nm to sub-20 nm diameter and 30 to 50 nm thickness, fabricated by electron beam lithography (EBL) and transmission electron microscopy (TEM) drilling. We experimentally investigate the scaling laws of *N*-pore conductance across several orders of magnitude in salt concentrations and the ion current rectification under salt gradient conditions. We also develop 3D model simulations based on Poisson-Nernst-Planck theory to provide insights into the underlying mechanisms of nonlinear ion transport in nanopore arrays. The 3D simulations involve a large system size, where the array contains up to 36 nanopores within a 50 nm membrane thickness. The theoretical results agree with the experimental observations including the sublinear increase of *N*-pore conductance and the reduction in current rectification in NPA. These behaviors can be explained by analyzing the ionic concentration and electric field profiles, showing the concentration polarization and the electric field overlapping in NPA. Finally, we assessed the performance of osmotic energy conversion in NPA system and showed that the power density drops significantly with increasing number of pores up to three orders of magnitude from 4 to ~ 10000 pores.

## RESULTS AND DISCUSSION

### *Fabrication of Nanopore Arrays*

**Figure 1a** schematically illustrates the fabrication workflow of nanopore arrays via combined EBL and reactive ion etching (for diameters larger than 15 nm), similar to previous studies.[29] The suspended SiN membrane on the silicon chip was coated with a ZEP520A resist. The NPA was then patterned using EBL onto the SiN membrane window. To achieve the minimum possible feature size, cold development[30] was employed at -10°C (**Figure S1**). Subsequently, the array pattern was transferred by using reactive ion etching (RIE) to remove the exposed region of SiN, thus creating cylindrical pores. Details of fabrication parameters are provided in the **Methods**. By tuning the electron irradiation dose in EBL, shot pitch and pattern size, the array configuration can be precisely controlled. **Figures 1b-e** and **S2** show TEM images of NPAs for various combinations of pore numbers ($N$) and the interpore distances ($L_{int}$), highlighting the flexibility and scalability of the EBL platform. TEM images also reveal that the nanopores are round and uniformly distributed and are characterized by diameters of ~ 20 nm (see **Methods**), demonstrating the reliability of the fabrication. We also observed that for NPAs having 20 nm diameter pores, the membranes maintain their structural integrity and do not collapse when the interpore distances exceed 40 nm.

For smaller diameter pores, TEM drilling was employed. **Figure 1f** shows NPAs with sub-3 nm diameter pores on SiN windows (see **Methods**). For these TEM-drilled NPAs, we focused on making 2 by 2 square arrays with various pore-to-pore distances spanning from ~15 nm to 150 nm, and the smallest pore spacing of ~ 13 nm (**Table S1**). The average pore diameter of the four NPAs ranged from 1.9 nm to 4.1 nm. **Figures 1g-h** showcase two NPAs with average pore diameters of 2.2 ± 0.5 and 2.6 ± 0.4 nm, and interpore distances of ~ 40 and ~ 20 nm, respectively. The robustness of the drilling technique was further confirmed by making and measuring over 10 devices, where pore diameter was mostly in the 2-3 nm range and smallest pore diameter was 1.4 nm (**Table S1**). Details and additional TEM images for these sub-3 nm four-pore arrays are provided in **Supplementary Table S1** and **Figure S3.**

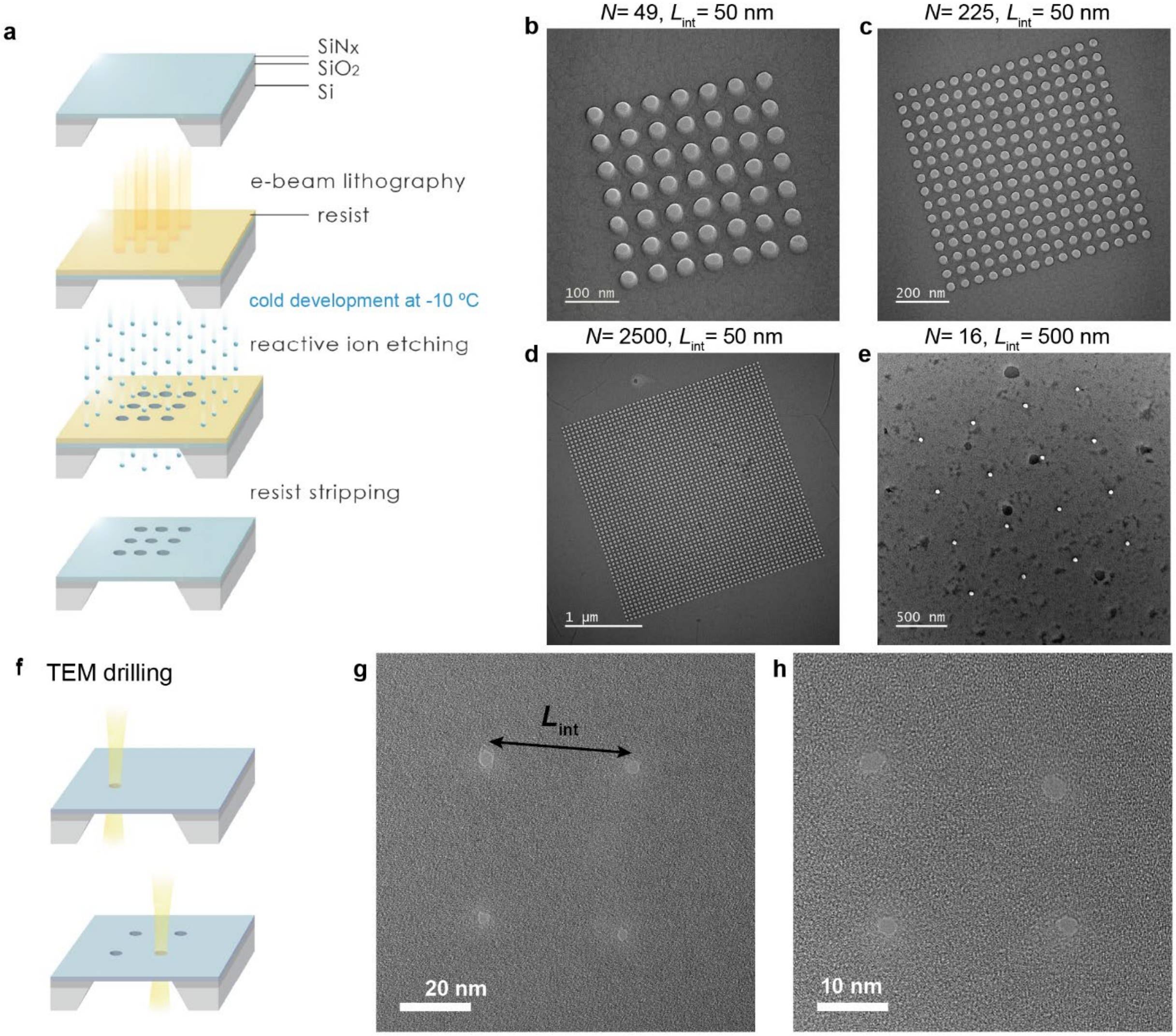

**Figure 1.** Two methods of nanopores array (NPA) fabrication. (a-e) Schematic illustration of EBL and RIE method, (a), and the associated TEM images, (b-e). (b) $N$ = 49 and $L_{int}$ = 50 nm; (c) $N$ = 225 and $L_{int}$ = 50 nm; (d) $N$ = 2500 and $L_{int}$ = 50 nm; (e) $N$ = 16 and $L_{int}$ = 500 nm. (f) Schematic illustration of TEM-drilling method for a sub-3 nm diameter square array. (g) Average pore diameter $d_{pore}$ = 2.2 ± 0.5 nm and interpore distance $L_{int}$ = 44 ± 3.0 nm. Membrane thickness is 30 nm. (h) Average pore diameter $d_{pore}$ = 2.6 ± 0.4 nm, and interpore distance $L_{int}$ = 23 ± 2.9 nm. Membrane thickness is 20 nm. The standard deviation is from the analysis of TEM images (see **Methods**).

***Scaling of Ionic Conductance***

To examine the scaling law of NPA conductance on the number of pores, $N$, we measure the conductance of the NPA, $G_N$, for various $N$ from 1 to 10 000, diameters ~ 2 nm to 20 nm, and $L_{int}$ = 13 to 1 μm, at 1M KCl (**Figure 2**). As shown in **Figure 2a**, $G_N$ increases with increasing $N$ and starts to deviate from the linear scaling when $N$ is sufficiently large (in this case $N$ ~ 49). Here, the gray line represents the linear scaling $G_N = N \times G_1$, which equals the number of pores times the conductance of a single pore. We assume that all the pores within the NPA are of the same size, which is roughly the case, by design.

In this work, to quantify the degree of deviation from this simple scaling relation, we normalized $G_N$ by dividing it by $N \times G_1$. For the ideal case of independent or non-interacting pores, the $N$-pore conductance would scale linearly with the number of pores, thus $G_N/N \times G_1 = 1$. As can be seen in the inset of **Figure 2a**, $G_N$ decreases to below 0.1 of its ideal value $N \times G_1$ when NPA contains ten thousand pores. On the other hand, for a fixed number of pores ($N$ = 100) and variable pore separation $L_{int}$, $G_N$ clearly increases with increasing $L_{int}$ from 31% of $N \times G_1$ to 87 % of $N \times G_1$ (**Figure S4**). This means that the interaction between individual pores is mitigated when the pores are far away from each other, which is qualitatively consistent with previous studies with small $N$ = 2 and 3.[12, 14, 15]

Taking into account the entrance effect (i.e., the conductance contributions from spatial regions above and below the pore), the $N$-pore dependent conductance can be described as:[12]

$$G_N = N\kappa_b\left[\frac{4L_\mathrm{p}}{\pi {d_\mathrm{pore}}^2} + \frac{1}{d_\mathrm{pore}}\left(1+\gamma\frac{d_\mathrm{pore}}{2L_\mathrm{int}}\right)\right]^{-1} \quad (1)$$

Here, $G_N$ is the conductance of the array, $N$ the number of pores, $d_{pore}$ the pore diameter, $L_p$ the effective pore thickness, and $L_{int}$ the interpore distance. $\kappa_b$ is the bulk ionic conductivity, and $\gamma$ is a factor accounting for the array configuration and the decay of conductance.[12] It should be noted that, for the case of a single pore ($N$ = 1), $\gamma$ is constrained to be 0, and **Eq. 1** reverts to the classical conductance model, $G_1 = \kappa_b\left[\frac{4L_\mathrm{p}}{\pi {d_\mathrm{pore}}^2} + \frac{1}{d_\mathrm{pore}}\right]^{-1}$. Besides these equations, other efforts to develop analytical models for the $N$-pore conductance have been reported in the recent literature and warrant attention.[31, 32]

To elucidate the underlying physics of transport behavior within the NPA, we also developed a 3D model by solving numerically Poisson-Nernst-Planck equations (see **Methods**).[26] Our modeling successfully reproduces the dependence of $G_N$ on $N$ (symbols of **Figure 2c**), that is, $G_N/N\times G_1$ appreciably decreases to 0.75 with increasing $N$ from 1 to 36. The simulation data is further compared with **Eq. 1**, with the best fitting obtained for $\gamma = N^{0.65}$ at 1 M KCl. This exponent value is larger than that predicted by Gadaleta et al.[12] for a square array ($\gamma \sim N^{0.5}$). This difference can be attributed to surface charge and concentration polarization effects that were not accounted for in their analysis and that lead to a more pronounced decrease in $G_N/N\times G_1$. In the case of lower concentrations, $G_N/N\times G_1$ decreases more strongly with increasing $N$, resulting in a larger fitting exponent $\gamma$, e.g., $\gamma = N^{1.15}$ at 1 mM KCl and $\gamma = N^{0.8}$ at 10 mM KCl (see **Figures 2d** and **S5**). This is not surprising because the concentration polarization effect becomes more pronounced at lower concentrations, which consequently hinders ion transport through the NPA membrane.[15, 25, 26] The result indicates that $\gamma$ is concentration-dependent, and that caution should be taken when applying **Eq. 1** to array system with surface charge and small scales.

Next, we examined the scaling of $G_N$ with $N$ for NPAs at different salt concentrations. **Figure 2b** presents the conductance ratio of NPAs with $N$ = 25 and $N$ = 4 (= $G_{N,25}/G_{N,4}$) for various KCl concentrations. If the conductance is ideally scaled up, the ratio would be 25/4 = 6.25 (gray line in **Figure 2b**). In general, the deviation from the scaling law remains evident even for smaller $N$. At 1 M KCl, the measured $G_{N,25}/G_{N,4}$ equals to 4.9, which matches well with the theoretical values from PNP theory (5.1, blue curve) and analytical formula (5.5, symbol). The difference between the theoretical and experimental results becomes larger at lower concentrations, but the qualitative trend remains the same. Interestingly, $G_{N,25}/G_{N,4}$ decreases as the concentration decreases from 100 mM to 1 mM, suggesting that lower concentrations amplify the deviation from the scaling behavior. This can be attributed to the stronger concentration polarization at lower concentrations where the surface-charge-governed regime dominates.

**Figures 2e** and **2h** show the axial variation of concentration and the corresponding 2D plot in the x-z plane for arrays with *N* = 1 and *N* = 25 at a bulk concentration of 1 mM KCl. When an external electric field is applied, cations migrate along the field direction, leading to the formation of an accumulation zone on one side of the pore and a depletion zone on the other side (**Figure 2h**). This concentration polarization leads to a stronger reversed electric field at the nanopore interface (i.e., access region), which hinders the transport of ions across the membrane (red curve of **Figure 2f**). Compared with *N* = 1, the *N* = 25 case exhibits enhanced ion enrichment on the grounded side (left side of **Figure 2e**, **Figure 2h**) and more pronounced ion depletion on the opposite side (right side of **Figure 2e**, **Figure 2h**). As a result, the larger the *N* and/or the lower the concentration, the more significant is the polarization of ion concentration. Note that a thicker diffusion layer outside the nanopore is observed for a larger NPA (*N* = 25), in line with previous theoretical analysis[25]. In addition, the magnitude of the electric field is reduced as *N* increases (**Figure S6**), which also explains why the conductance does not increase linearly with *N*. We further plotted the 2D profile of ionic concentrations in the x-y plane on the two sides of nanopore entrances (**Figures 2i-k**). **Figure 2i** clearly shows that both accumulation (z = 0 nm, bottom) and depletion (z = 30 nm, top) zones of ions are overlapping. This phenomenon becomes pronounced as *N* increases (**Figures 2j**) and NPA becomes more packed (**Figures 2k**).

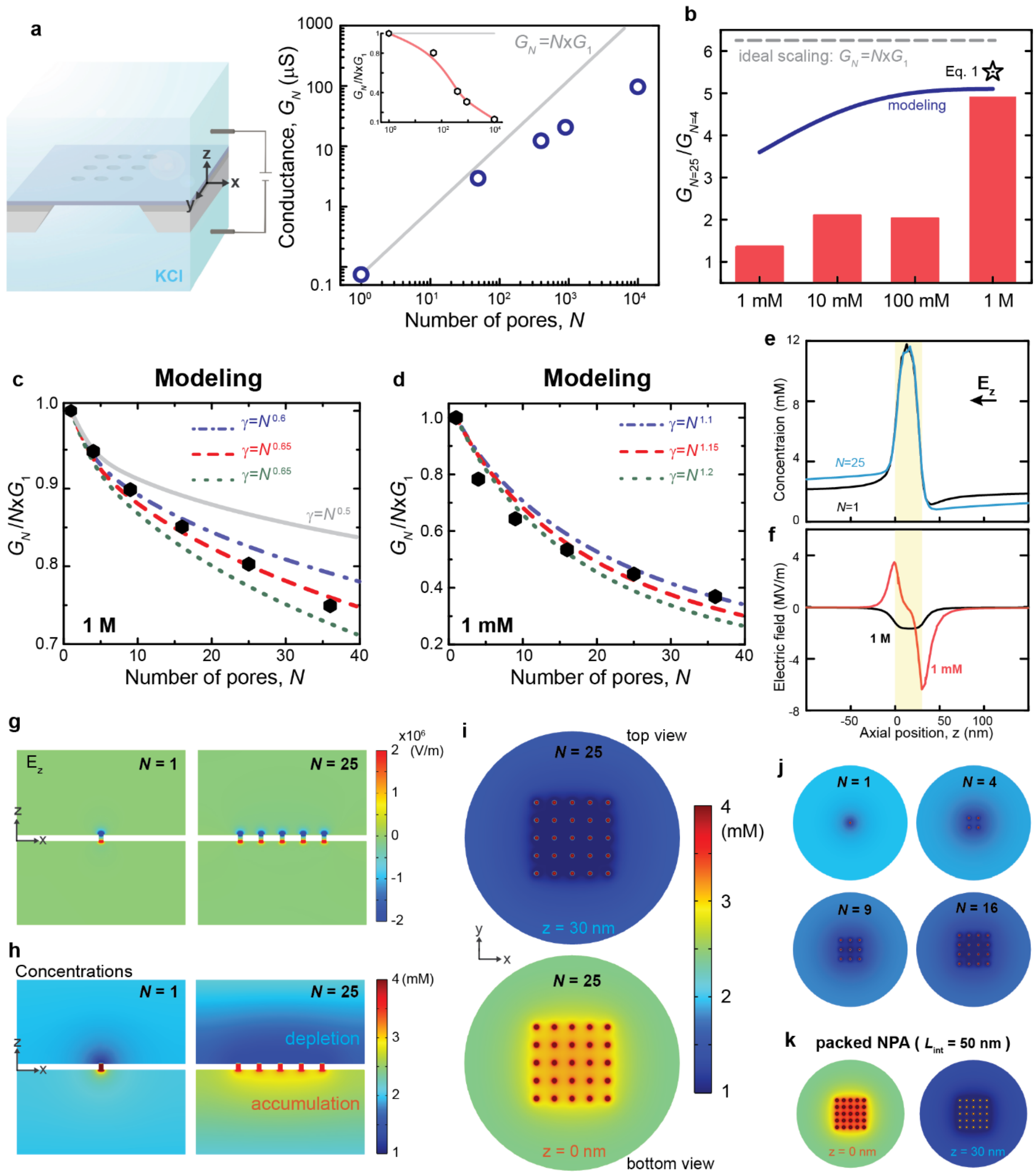

**Figure 2**. *Ionic measurement*: ionic conductance in NPAs in KCl solution, $N$ = 1 to 10 000. (a) Schematic of the ionic measurement with NPA and the measured ionic conductance, $G_N$, as a function of number of pores, $N$, at 1 M KCl. Gray line represents the linear scaling relation, $G_N = N \times G_1$. Inset: the corresponding normalized conductance $G_N/(N \times G_1)$ as a function of $N$. $G_1$ denotes the conductance of one pore ($N$ = 1). (b) Conductance ratio between NPAs with $N$ = 25 and $N$ = 4 for various KCl concentrations from 1 mM to 1 M. Blue curve: 3D model simulation. Symbol: analytic formula **Eq. 1** with $\gamma \sim N^{0.65}$. *Theoretical*

*modeling*: (c, d) Modeling of normalized conductance $G_N/(N\times G_1)$ versus $N$ for two different concentrations, 1 M KCl, (c), and 1 mM KCl, (d). Dashed curves: results obtained from **Eq. 1** with various fitting of $\gamma$. Gray curve in (c) represents the result using the proposed parameter (i.e., $\gamma \sim N^{0.5}$) in the literature[12]. (e) Axial variation of ionic concentration along z-axis for two different $N$ = 1 and $N$ = 25 at 1mM KCl. (f) Axial variation of electric field along z-axis for a NPA with $N$ = 25 at 1 M (black curve) and 1 mM (red curve) concentrations. Yellow shaded area denotes the nanopore interior region. z = 0 and 30 nm define the x-y planes corresponding to the pore entrances, on the grounded and applied-voltage sides, respectively. (g, h) 2D profiles of electric field, (g), and ionic concentrations, (h), in the x-z plane of the 3D model for cases of $N$ = 1 and $N$ = 25 at 1 mM KCl. (i-k) 2D profiles of ionic concentrations in the x–y planes of the 3D model for cases of varying $N$ and *Lint* at 1 mM KCl. z = 0 nm and z = 30 nm planes correspond to the x–y planes at the pore entrance on the grounded and applied-voltage sides, respectively. (i) $N$ = 25 and *Lint* = 100 nm. (j) $L_{int}$ = 100 nm is fixed by varying $N$ (= 1, 4, 9, 16). (k) $N$ = 25 and $L_{int}$ = 50 nm.

### *Evolution of Conductance-Concentration Scaling Behaviors*

Furthermore, we measured the conductance of the NPA, $G_N$, as a function of KCl concentrations for different combinations of $N$ and $L_{int}$ (**Figure 3**). It has been established that transport of ions under low salt concentrations is dominated by the surface charges of nanopore walls, resulting in an enhanced conductance and even a conductance plateau regime at low concentrations.[33, 34] Conversely, this conductance behavior at low concentrations has been widely adopted as a proof of the existence of surface charges in nanofluidic devices,[35, 36] nanopores,[37-40] biological ion channels,[41, 42] and nanoporous membranes.

We show that the saturation of $G_N$ at low concentration also occurs for NPAs with small numbers of pores but disappears as $N$ increases. In **Figure 3a** we show $G_N$ vs. concentration on a log-log plot for $N$ = 4. $G_N$ departs from the bulk-conductance behavior and reaches a plateau at lower concentrations, indicating the effects of surface charge of SiN. This characteristic two-regime curve is well described by a linear scaling of a conductance of single pores that incorporates surface charge effects (**Eq. 2**, **Methods**), yielding a fitted charge density of -20 mC/m$^2$. This value is consistent with reported values for SiN nanopores.[43] However, as $N$ increases, the surface-charge-governed behavior gradually disappears (see **Figures 3b,c**). While $G_N$ still deviates from the linear scaling at the lower concentration range, the deviation is less significant; in other words, the surface-charge-governed ion transport becomes less pronounced. Note that the gray lines are the linear extrapolations of the conductance recorded at 1 M KCl, for comparison. This can be inferred from our observation of the concentration-dependent scaling behavior shown in **Figure 2**. At a low concentration, the conductance scaling is more suppressed with increasing $N$, leading to a lower $G_N$, and therefore, approaching the linear bulk conductance regime. This observation also reveals that the contribution of surface charge becomes less significant and NPA tends to behave like a less selective membrane when $N$ is sufficiently large, in contrast to the widely observed conductance plateau for single pores. *We want to emphasize that this does not mean a loss of surface charges along the nanopore wall, as the intrinsic selectivity remains intact due to the nature of the material (i.e., silanol and amine groups).* Rather, the apparent selectivity is much lower than the intrinsic selectivity of NPA. This is also qualitatively consistent with a recent theoretical study showing that the apparent selectivity of the membrane rapidly declines when the pore density is high enough, which is attributed to the concentration polarization at the membrane interface.[25] Furthermore, we applied the **Eq. 2** (see **Methods**) on the $G_N$ vs concentration curve obtained from the NPA with $L_{\mathrm{int}}$ = 1000 nm (**Figure 3c**). Since the individual pores are far away from each other, the suppression of $N$-pore conductance is mitigated, thus the conductance at 1 M fits better with **Eq. 2** even if $N$ = 100. **Figure 3c** clearly shows that the extracted surface charge density (~1 mC/m$^2$) fits well with the experimental data, which is significantly lower than the typical value reported for SiN (20 mC/m$^2$)**.** Such a reduced charge density is consistent with those reported in the literature, where conductance-based models (similar to **Eq. 2**) were used to fit the conductance of multipore and multichannel systems to extract the membrane charge density.[44, 45] *Our finding not only highlights the limitation of simple analytical models but also alerts researchers that surface charge densities extracted from fitting multiporous membranes using such approaches may not accurately describe the intrinsic properties of the membrane.*

To gain further insight into the regime of NPAs with small pore diameters, we examine the conductance vs. concentration scaling behavior with sub-3 nm diameter NPAs (**Figures 3d-f**), which are fabricated by TEM-drilled method (**Figure 1** and **Methods**). **Figures 3g-i** include the corresponding TEM images displaying square arrays with various interpore distances from ~ 25 nm to ~ 150 nm, corresponding to $L_{int}/d_{pore}$ from ~10 to 53. When the nanopore size becomes small, the individual pores may behave more independently when they are separately far enough. In this scenario, the total conductance at a high concentration (e.g., 1M KCl) follows the linear superposition of the individual nanopore conductances, as demonstrated in our previous work with two 3 nm diameter nanopores separated by 1 µm apart.[14] As shown in **Figure 3**, the measured conductances of the sub-3 nm diameter NPAs ($N$ = 4) at 1 M KCl are in good agreement with the values from **Eq. 1** based on the TEM-measured pore diameters and effective thickness (**Methods**), even when pore separation is small ($L_{int}$ = 24 nm, **Figure 3d**). This is likely attributed to the thin electrical double layer (EDL) at 1 M KCl (Debye screening length ~ 0.3 nm), which renders pore interactions negligible at these interpore separations. Note that the conductance of a smaller diameter ($d_{pore}$ = 2.3 nm) and closely packed ($L_{int}$ = 24 nm) NPA (e.g., 1M KCl in **Figure 3d**) is slightly larger than that of a larger diameter ($d_{pore}$ = 3 nm) and loosely packed ($L_{int}$ =74 nm) NPA (**Figure 3e**). The effective thickness $L_p$, derived from **Eq. 1**, are 10 nm (**Figure 3d**) and 30 nm (**Figure 3e**), respectively. This may result from (i) the smaller effective thickness when fabricating packed NPA by TEM-drilling or (ii) incomplete wetting of all nanopores. In such closely packed arrays, the drilling process is confined to a small area, and this confined region is simultaneously irradiated by the electron beam, leading to progressive thinning of the material across the entire region (**Methods**). Operating in STEM-drilling may help address this issue by using smaller probe size and thereby enabling the fabrication of more densely packed NPAs. Such development can be pursued in future work.

Interestingly, anomalous ion transport behavior was observed upon spanning the concentration from high to low. As shown in **Figure 3d**, when NPA is densely packed ($L_{int}$ = 24 nm, **Figure 3g**), $G_N$ vs. concentration curve exhibits an approximately straight line on a log-log scale, and the typical saturation of conductance at lower concentrations is not observed. However, if compared it to the bulk conductance (gray line), the measured conductance $G_N$ exhibits significant deviation as bulk concentration is below 100 mM (see **Figures 3d,e**). In contrast, when the nanopores are sufficiently far apart ($L_{int}$ = 143 nm), $G_N$ vs. concentration curve gradually evolves into a two-regime-like curve, typically characteristic of single pores[37] (**Figure 3f**). Conventionally, a deviation in conductance from bulk conductance at low concentrations has been taken as evidence of surface charge in nanofluidic devices, while a linear conductance vs. concentration relationship is interpreted as indicative of a neutral or weakly charged surface.[35, 41] Although this criterion may be valid for single nanopores or nanochannels, our result reveals that it can be misleading in multipore or porous systems, where linearity-based analysis is insufficient for reliable interpretation. *In other words, in multipore systems, the surface charge density extracted from conductance-based measurements may be underestimated and does not fully capture the intrinsic properties observed in single nanopore systems.*

The disappearance of a two-regime characteristic curve has been reported in single carbon nanotubes,[36, 46] where the conductance was found to follow the power law dependence as a function of concentration ($G \sim c^{\alpha}$) with exponent ($\alpha$) close to 1/3. The physisorption of hydroxide ions on the carbon surface was proposed to account for this behavior.[36] In the case $\alpha = 1$, the conductance has a linear relationship with concentration, whereas for $\alpha = 0$, it saturates with concentration. The dependence of $\alpha$ on pore length, pore diameter, and surface charges has been theoretically discussed with single nanopore in a recent theoretical work.[47] In contrast, our system involves multiple interacting nanopores, making the transport behavior more complex, and to our knowledge, such behavior has not been reported in NPAs. We performed the power law fitting for our sub-3 nm diameter NPAs and obtained $\alpha = 0.55$ and $\alpha = 0.53$ for NPAs with $L_{\mathrm{int}}$ = 24 nm (**Figure 3d**) and $L_{\mathrm{int}}$ = 75 nm (**Figure 3e**), respectively. The larger value of $\alpha$ indicates that our system exhibits behavior closer to linear bulk behavior. We attributed it to the strong pore interaction when nanopores are closely packed.

We also compared the measured $G_N$ vs. concentration with values obtained from **Eq. 2** by using typical values of surface charge density for SiN nanopores (**Figures 3d-f**). The dashed curves can be regarded as the ideal scaling case with different surface charge densities, in which the contribution from surface charge is also incorporated as a linear addition. It is evident that at low concentrations, although the measured $G_N$ still deviates from the bulk conductance (gray line), it remains significantly lower than the ideal scaling case when assuming surface charge density of 20 mC/m$^2$. A lower charge density (blue curves) may capture more experimental data points but still fails to provide a good fit at low concentrations. This observation suggests that, in the low-concentration scenario, even for a relatively small number of nanopores ($N$ = 4) and small diameters (sub-3 nm), strong pore-to-pore interactions can still manifest when nanopores are closely spaced (i.e., small $L_{int}$), leading to strong suppression of conductance at low concentrations. Consequently, charge densities estimated from such $G_N$ vs. concentration curves show an apparent decrease.

We also want to note that analytical equations inherently become less accurate for small pore diameters. The sub-3 nm diameter in our NPA pushes the system close to the validity limits of these analytical models, since at this scale, the finite size of ions and the van der Waal forces also become increasingly important.[47, 48] In addition, the constant charge density is unable to account for the pH-dependent behavior of the SiN surface[49, 50] and therefore overestimates the contribution of surface charges to conductance in the low concentration regime.[36, 37]

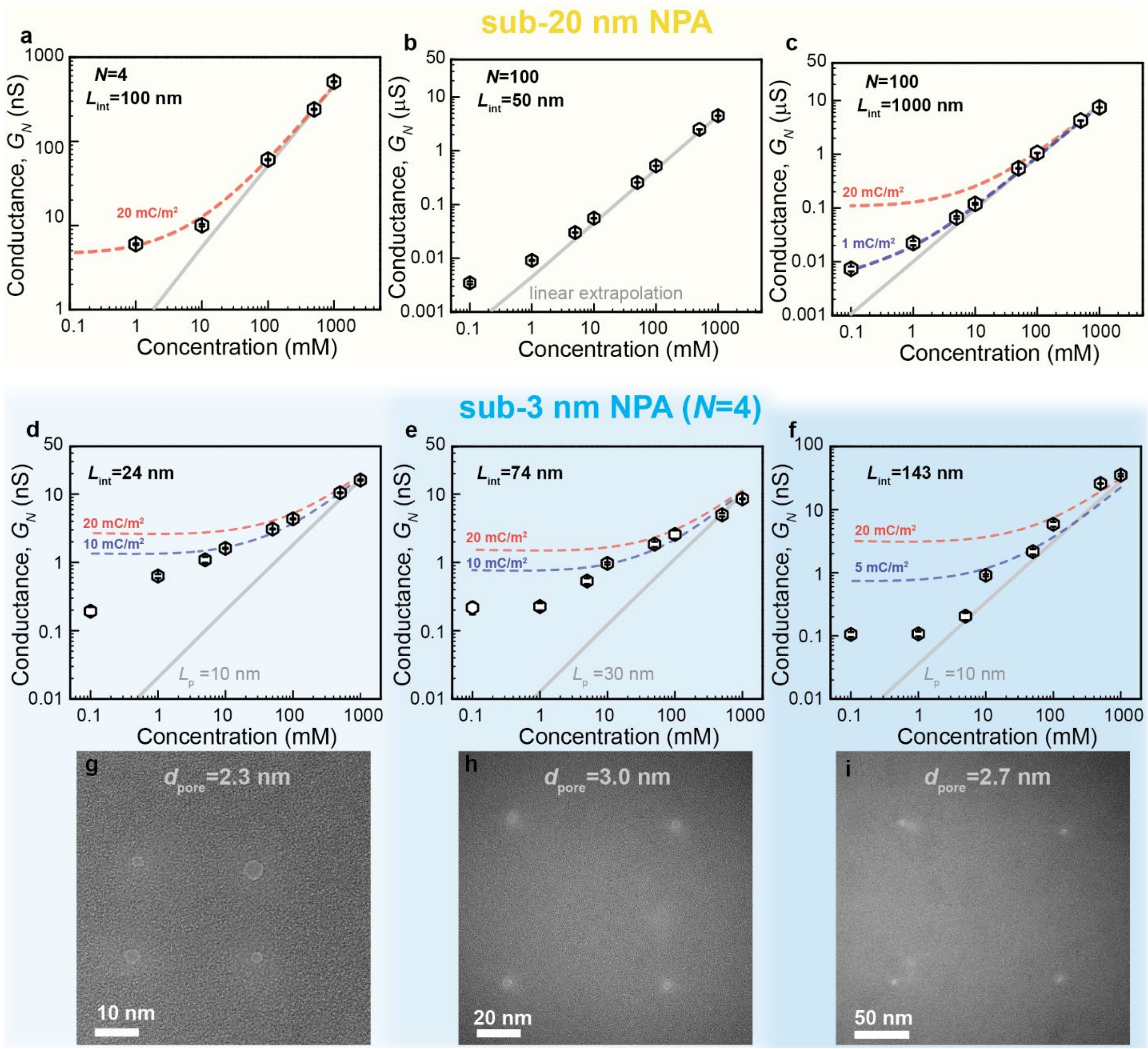

**Figure 3.** *Ionic measurement*: Scaling law of ionic conductance of NPA with salt (KCl) concentrations. (a-c) $G_N$ versus concentration for sub-20 nm diameter NPAs for various pore number, $N$ = 4 to 100, and interpore distance, $L_{int}$. (a) $N$ = 4 *and* $L_{int}$ = 100 nm; (b) $N$ = 100 *and* $L_{int}$ = 100 nm; (c) $N$ = 100 *and* $L_{int}$ = 1000 nm. Gray line in (a-c) denotes the linear extrapolations of the conductance recorded at 1 M KCl. (d-f) $G_N$ versus concentration sub-3 nm diameter NPAs with different interpore distances, $L_{int}$, and the corresponding TEM images below the graphs, (g-f). (g) $d_{pore}$ = 2.3 ± 0.3 nm and $L_{int}$ = 24 ± 0.3 nm; (h) $d_{pore}$ = 3.0 ± 0.4 nm and $L_{int}$ = 74 ± 0.8 nm; (i) $d_{pore}$ = 2.7 ± 0.4 nm and $L_{int}$ = 143 ± 5.4 nm. Dashed curves represent the results fitted by using the linear scaling of **Eq. 2** (**Methods**). Gray line denotes the linear scaling of the conductance without considering surface charge, $N*G_1$. Membrane thickness is 30 nm.

***Ion Current Rectification***

Salt gradient across the interior and exterior of the living cell membrane is essential to regulate transmembrane potential and ion current.[51] Therefore, understanding the transport phenomena under a salt gradient, such as current rectification and osmotic energy, is of great importance. Ion current rectification (ICR) in nanopores and nanofluidics has been extensively studied in the past two decades. Typically, ICR in a charged nanopore results from the breakdown of symmetry in geometry,[52] surface charge distribution,[53-55] salt concentrations,[50] wettability,[56] or a combination of these factors.

**Figure 4** shows the typical current-voltage curves of NPAs with $N$ = 25, 49, 900 and 1000 under salt gradient of 100 mM/1mM, where we show how the I-V curves are nonlinear and asymmetric for small $N$, and become linear as $N$ increases. **Figure 4a** shows the I-V curve for $N$ = 25 and $L_{int}$ = 50 nm, where the asymmetry and current rectification are observed. Additional ICR measurements from different NPAs are presented in **Figure S7**. This rectification effect can be explained by considering the nanopore charge. Specifically, in our case, nanopores are cylindrical and symmetric[29] and at pH 8, the SiN surface is negatively charged so the cations (potassium ions) dominating ion transport in the nanopore interior. The voltage was applied over the NPA membrane with the ground electrode placed in the flow cell having the lower KCl concentration. Due to the salt concentration difference across the membrane, ions are driven by the concentration gradient from the high concentration side toward the low concentration side. When a positive voltage is applied, the migration and diffusion ionic fluxes are in the same direction from high concentration side to low concentration side, resulting in an enrichment of ions inside the nanopores and, consequently, a larger ion current. In contrast, when a negative voltage is applied, the migration flux opposes the diffusion flux, leading to a smaller ion current. As a result, the ICR phenomenon is observed.

For larger $N$ = 49, 900, 10000, the current-voltage characteristics become gradually more linear as $N$ increases to 10000 (**Figures 4b-d**), suggesting that the ICR phenomenon gradually vanishes. This implies that the membrane appears to no longer exhibit ion selectivity at sufficiently large $N$. We then quantify the degree of I-V asymmetry by introducing the rectification factor $R_f$, which is defined as the absolute ratio of the current at a given voltage to that at the opposite polarity. This is a somewhat arbitrary measure because it depends on the value of the voltage chosen. Here, we compare currents at the maximum voltage used V = ± 1 V and define $R_f$ as $R_f=|I(+1V)/I(-1V)|$ ($R_f=1$ means no rectification). **Figure 4e** shows $R_f$ as a function of $N$ (for the data in **Figures 4a** to **4d**). Here the interpore distance was fixed to $L_{int}$ = 50 nm for all four arrays and we measured $G_N$ at a 100-fold salt gradient (100mM/1mM). We show that $R_f$ significantly decreases as $N$ increases from ~ 3 to 1.2, and $R_f$ is approximately unity for a NPA having $N$=10000. This means that while salt-gradient induced ion transport and rectification effects are observed in single pores, for large nanopore arrays, the rectification is lost. The rectification phenomenon progressively weakens upon scaling to larger $N$.

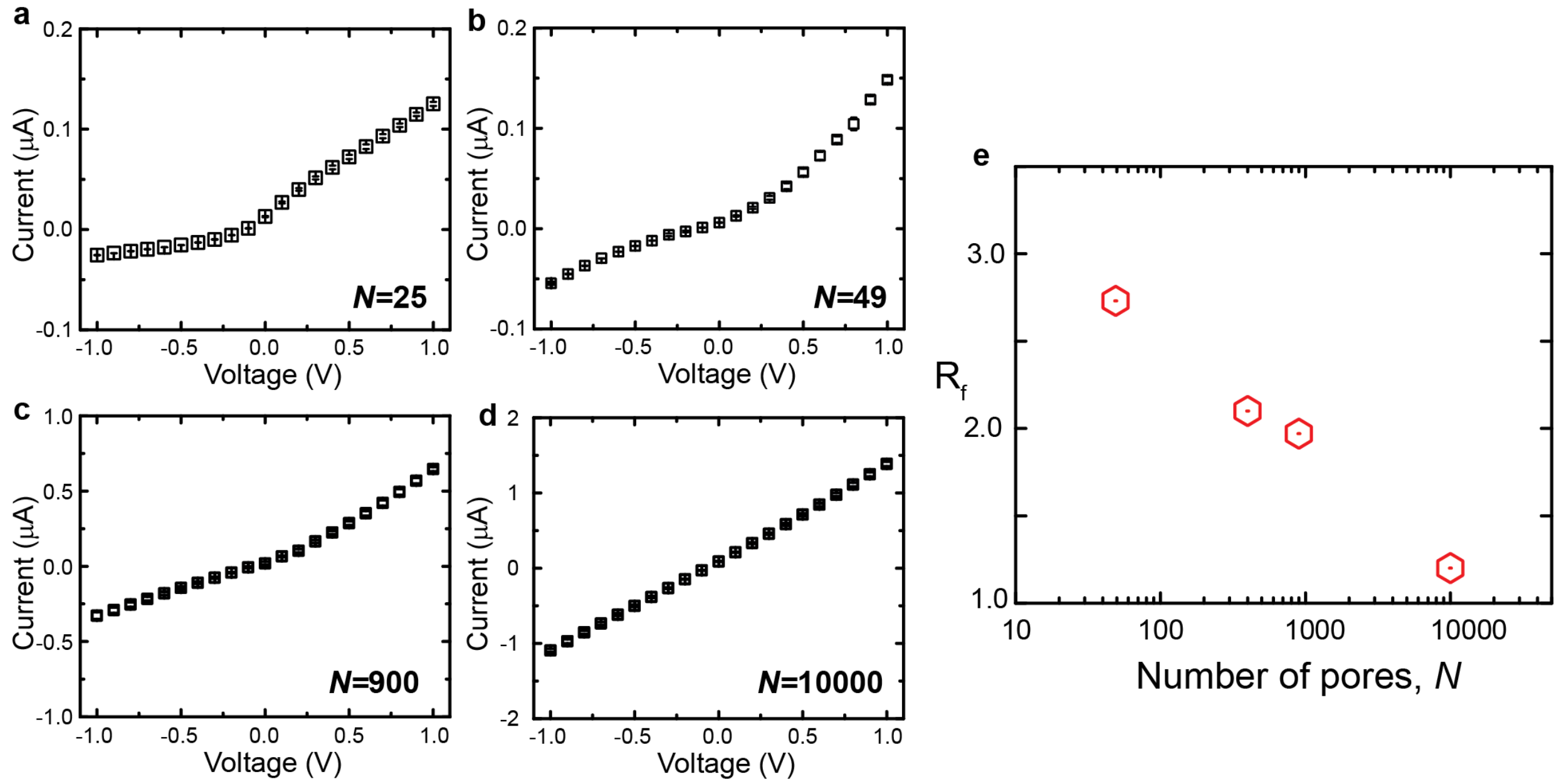

**Figure 4**. *Ionic measurement*: Ion current rectification with the presence of salt gradient in NPA. (a-d) Current-voltage characteristics of NPAs for different $N$ at a salt gradient of 100 mM/1 mM. (a) $N$ = 25 and $L_{int}$ = 100 nm; (b) $N$ = 49 and $L_{int}$ = 50 nm; (c) $N$ = 900 and $L_{int}$ = 50 nm; (d) $N$ = 10000 and $L_{int}$ = 50 nm. (e) Rectification factor, $R_f$, vs. number of pores, $N$, on a linear-log scale. $R_f$ is defined as $|I(+1V)/I(-1V)|$. $L_{int}$ is fixed at 50 nm.

**Figure 5** shows theoretical results obtained by modeling the I-V curves for NPAs with $N$ = 4 and $N$ = 36 for $L_{int}$ from 25 nm to 500 nm, and $R_f$ as a function of $N$ and $L_{int}$. The pore diameter and thickness are fixed at 20 nm and 30 nm, similar as the experiments.

Thus, the simulations are performed over an $L_{int}/d_{pore}$ range of 1.5 to 25. **Figure 5a, b** shows the current-voltage curves at a salt gradient of 100mM/1mM for various $L_{int}$ and $N$. For $N$ = 4, the simulated current-voltage curves show little difference for varying $L_{int}$ (**Figure 5a**). As $N$ increases to 36, the difference between I-V curves from closely packed ($L_{int}$ =25 nm, black curve) and loosely packed ($L_{int}$ =200 nm, red curve) NPAs is clearly observed, where the ion current of closely packed NPA is less rectified (**Figure 5b**).

The theoretical rectification factor $R_f$, calculated from the simulated I-V curves of NPAs, as a function of $N$ (= 1, 2, 4, 9, 16, 25, 36) is plotted in **Figure 5c** and as a function of $L_{int}$ in **Figure 5d** for $N$ = 2, 4, 9 and 36. As shown in **Figure 5c,** the rectification factor decreases with increasing $N$, which agrees well with the experimental observation (**Figure 4e**) that NPA having larger $N$ displays weaker rectification. This can be explained by the significant concentration polarization occurring in NPAs with a large number of pores, as supported by the variation in concentration profiles (**Figure S9**). The concentration at the membrane interface on the high concentration side is reduced when $N$ increases from a single ($N$ = 1) to multiple pores ($N$ = 25). On the other hand, on the low concentration side, the concentration for $N$ = 25 is higher than that for $N$ = 1. Therefore, the effective salt gradient across the NPA membrane is reduced with increasing $N$ and the driving force due to the salt gradient is weakened accordingly. The phenomenon of concentration polarization has also been employed to explain the huge reduction in osmotic energy in a multiple pore system.[25] **Figure 5d** shows that the rectification ratio decreases with decreasing $L_{int}$, with the reduction becoming more significant for NPAs with larger $N$. When the NPA contains a small number of pores (e.g., $N$ = 2), the rectification ratio almost remains the constant regardless of interpore separation ($L_{int}/d_{pore}$ ranges from 1.25 to 25). The result indicates that the ability of NPA to rectify current may partially recover when individual nanopores are located far away from each other. We also conducted additional simulations for thicker NPA membranes of 50 nm, as presented in **Figure S8**. The results exhibit a similar trend qualitatively, with a slightly higher magnitude of $R_f$, which can be attributed to the higher selectivity of thicker membrane.

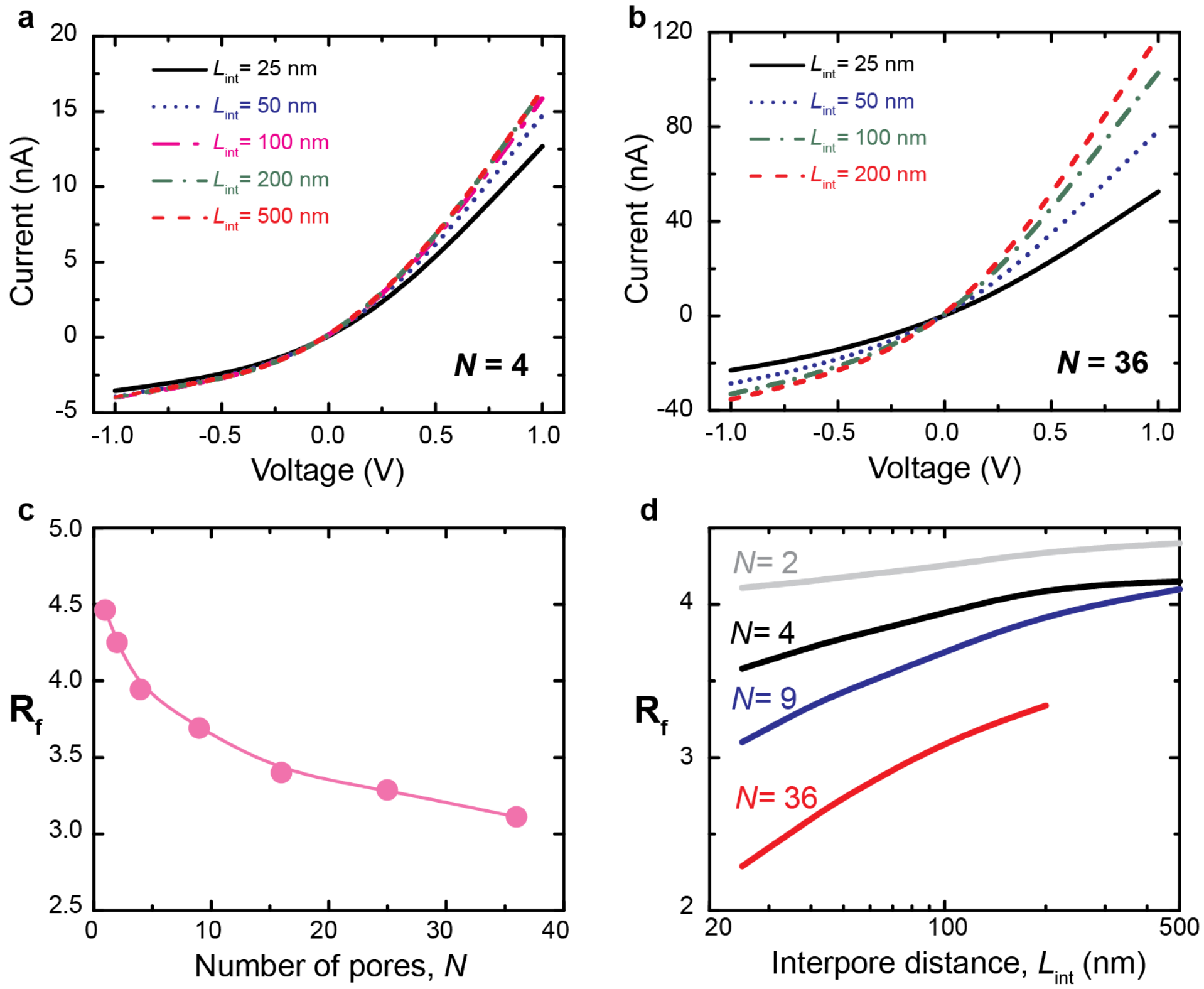

**Figure 5.** *Theoretical modeling*: 3D simulation of salt-gradient-driven ion transport in nanopore array. (a, b) Simulated current-voltage characteristics of NPA for various values of interpore distance, $L_{int}$, for $N$ = 4, (a), and $N$ = 36, (b), under a salt gradient of 100mM/1mM. (c) Variation of the rectification factor, $R_f$, with the number of pores, $N$, at a fixed $L_{int}$ = 100 nm. (d) Variation of the rectification factor, $R_f$, with $L_{int}$ for different values of $N$ = 2, 4, 9, 36. In (c) and (d), $R_f$ is defined as |I(+1V)/I(-1V)| and the salt gradient is fixed at 100 mM/1 mM. The NPA is comprised of nanopores with 20 nm in diameter and 30 nm in thickness.

### *Osmotic Energy Conversion*

The giant reduction of power density in multiple systems and its viability has been actively debated recently.[25, 28] To assess the NPA performance for applications in power generation, we also measured the osmotic energy by varying $N$ at a salt gradient of 100 mM/ 1 mM. By measuring the osmotic current ($I_{osm}$) and diffusion potential ($V_{diff}$), the maximum power density can be obtained by maximum output power ($P_{max}$) divided by the membrane area. An example I-V measurement on a SiN array with $N$ = 100 and $L_{int}$ = 50 nm is shown in **Figure 6a**. Choosing the membrane area is critical when interpreting the power density. A single pore system can achieve an extremely high-power density because of its small pore area. For comparison with our data, examples from the literature on the measurement of a single $MoS_2$ pore,[57] and single hBN nanotubes[58, 59] are shown as star symbols in **Figure 6b** where $N = 1$. Here, we adopt two reasonable methods to calculate area: the total cross-sectional area of the nanopores, $N^*\pi(d_{pore}/2)^2$, and the total area of the array, $N^*L_{int}^2$ (see the schematic in the right panel of **Figure 6b**). As shown in **Figure 6b**, the power density decreases dramatically with increasing $N$. Upon scaling to $N = 10000$, the power density decreases by about three orders of magnitude. By adopting the pore area, the power density can reach to ~$10^5$ W/m$^2$ for a NPA with $N = 4$. This value is comparable to the performance in single pore systems (see symbols in **Figure 6b** for $N = 1$), even though our platform features larger pore diameter (20 nm *vs.* 2-4 nm)[57, 58] and relatively lower surface charge (0.02 C/m$^2$ *vs.* 1 C/m$^2$)[59]. We also note that the membrane area considered here remains much smaller than that used in industrial ion exchange membranes (IEMs), which are typically on the cm$^2$ scale. For instance, the largest array of $N = 10000$ and $L_{int} = 50$ nm corresponds to a membrane area of 5 by 5 μm$^2$. Therefore, the obtained power density in our measurements appears to be higher than commercial benchmark (~1 to 5 W/m$^2$) of IEMs for reverse electrodialysis[60, 61]. Although material coatings on the SiN NPA may improve power density performance by applying gate voltage and changing surface charge distribution, existing studies are still limited to smaller $N$ ($N < 100$) and experiments performed on membranes with μm-scale areas.[62, 63] If scaled to industrial dimensions, the power density would be even reduced by several orders of magnitude in cm-scale membranes.

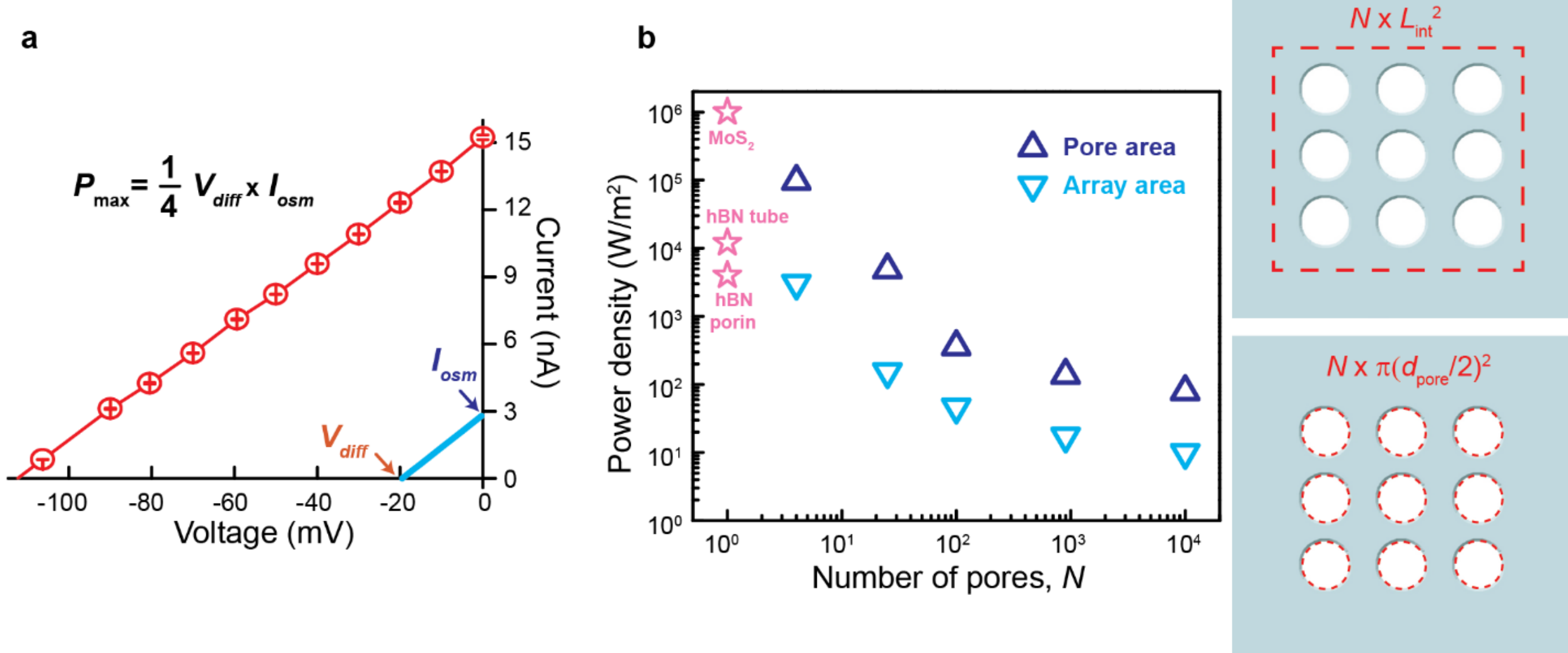

**Figure 6.** *Harvesting osmotic energy from sub-20 nm SiN NPAs*: (a) Osmotic energy conversion in solid-state SiN arrays. Schematic illustration of osmotic power measurement at a 100 mM/1 mM KCl. $I_{osm}$ and $V_{diff}$ are osmotic current and diffusion potential, respectively. $V_{diff}$ is obtained by subtracting the redox potential (in this case, 107 mV). $P_{max}$ is maximum output power calculated from $(I_{osm}*V_{diff})/4$. (b) Variation of power density with $N$ at a salt gradient of 100 mM/ 1 mM. $L_{int}$ = 50 nm and membrane thickness is 30 nm. As illustrated, the up and down triangles represent the membrane area using the total area of nanopores (= $N*\pi(d_{pore}/2)^2$) and the area of array (= $N*L_{int}^2$), respectively. Star symbols denote the single $MoS_2$ nanopore[57] and hBN nanotubes[58, 59] data from previous studies.

## CONCLUSIONS

In summary, we investigated the scaling behaviors in silicon nitride nanopore arrays by measuring ionic conductance, ion current rectification, and osmotic power, bridging fundamental ion transport physics and blue energy harvesting applications. We selected silicon nitride as the model material for studying scaling laws and geometric effects because silicon nitride pores are well characterized in the literature, including their composition, surface charge, and related properties. **Table 1** summarized results from prior studies and provided a comprehensive overview of SiN nanopore array measurements, helping to contextualize our findings and highlight broader trends across the literature.

We fabricated and studied both small, sub-3 nm-diameter nanopore arrays and large-scale ($N$ ~ 10000) sub-20 nm diameter nanopore arrays. Our experimental and modeling results confirm that the ionic conductance deviates from linear scaling as the number of pores increases, and that this deviation becomes more pronounced at lower concentrations. While the previous studies showed a similar trend on larger pore diameter (200 nm) and smaller number of pores ($N$ < 50),[12] here we present a cumulative and systematic study with $N$ varying by several orders of magnitude. Furthermore, we explain how this nonlinear scaling of conductance with $N$ can be explained by the concentration polarization at the nanopores entrance and the reduced electric field within the nanopore arrays, which collectively hinders the ion transport across the array membrane.

Another noteworthy aspect of our measurements includes uncovering how the conductance of NPA, $G_N$, depends on concentration with different array configurations. Understanding $G_N$ vs. concentration is an active topic of study even for single pores, especially in smaller diameter pores and pores of different materials.[35-38, 47, 64] Uncovering the collective $G_N$ vs. concentration behavior in arrays and linking it to the individual nanopore behavior has yet to be clarified. For sub-20 nm diameter arrays, the result shows that surface-charge-governed ion transport behavior is still observed at small $N$, but progressively weakens with increasing $N$ due to the strong pore interactions, resulting in a lower estimated charge density. For sub-3 nm diameter arrays, the conductance displays an anomalous dependence on concentration that follows a power-law relationship when the interpore distance is sufficiently small ($L_{int}$ ~ 25 nm). We also discussed effect of number of pores on the current rectification phenomenon by applying a salt gradient across the arrays. As $N$ increases, the I-V characteristic eventually approaches linear behavior and the associated rectification factor decreases. The rectification vanishes when $N$ is increased to a large number (10000). As the final step, the performance of osmotic power harvesting with sub-20 nm diameter SiN nanopore arrays is assessed. The power density significantly decreases with increasing $N$ and can decrease by up to three orders of magnitude when scaled to $N$ = 10000.

*We conclude that ion transport behavior in solid-state nanopore arrays differs fundamentally from that in single nanopores: many characteristics observed in single nanopores are significantly weakened or even vanish upon upscaling to arrays.* In addition, linear scaling of conductance on $N$ does not hold for large nanopore arrays. This deviation affects the transport phenomenon of ions and has a strong impact on osmotic energy conversion. Our results not only highlight the transport phenomena in nanopore arrays but also underscore the need for caution among researchers when using conductance-based measurements to quantify material properties in multipore systems.

## MATERIALS AND METHODS

*Fabrication of Nanopore Arrays.* Suspended SiN window is made through a traditional microfabrication method. NPA were fabricated via two approaches: (i) electron beam lithography and reactive ion etching; and (ii) TEM drilling. We utilized EBL and RIE approach to fabricate sub-20 nm diameter array.[10, 29] As illustrated in **Figure 1a**, SiN chip was first spin-coated using diluted ZEP520A followed by 3 minutes pre-baking at 180°C, yielding a film thickness of approximately 70 nm. EBL (Elionix 7500EX, 50 kV and 100 pA) was used to expose NPA pattern onto the SiN membrane window. NPA geometrical parameters, such as pore diameter and interpore distance, is optimized based on the dose and shot pitch of the pattern. The pattern is then developed in o-Xylene at -10°C for 70 seconds and rinsed with isopropyl alcohol (IPA) for 30 seconds. After cold development, RIE (Oxford 80 plus) with trifluoromethane and oxygen ($CHF_3/O_2$) was applied at an etching rate of ~ 1 nm/s to form NPAs on the SiN membrane window. To strip off the resist, the chip was treated with UVO for 10 mins and then immersed in heated N-methyl-2-pyrrolidone (NMP) at 70°C for 5 hours and finally rinsed with IPA.

For sub-3 nm diameter NPA, a high-resolution TEM (F200, JEOL Ltd.) operating at 200 kV was used to drill the nanopore arrays. Electron beam with emission current of 15 μA was focused at 1Mx magnification when performing in-situ drilling (TEM mode). Once the first pore was formed, either the beam was shifted or the stage was moved to the next desired location for drilling the subsequent pores. This procedure was repeated until four pores were fabricated (**Figure 1f**). We have successfully drilled NPAs on the SiN membranes with 20 and 30 nm thickness. Fabricating closely packed array using the TEM mode of drilling becomes challenging when pore-to-pore distance is below 20 nm because nanopores may merge or the neighboring nanopore may expand or shrink during the drilling process (**Figure S3d-f**).

*Characterization of pore diameter and thickness.* Nanopore diameters, $d_{pore}$, are determined by TEM images, corresponding to the smallest constriction region within the nanopore. **Figure S10** illustrates characterization of nanopore diameter. The membrane thickness is defined by ellipsometry. For NPAs fabricated by EBL and RIE approach, the nanopores show cylindrical geometry, thus the pore length is assumed as the membrane thickness.[29] The interpore distance is adopted from the shot pitch, which shows excellent agreement with TEM images. On the other hand, for TEM-drilled NPAs, interpore distance is defined as the center-to-center distance between individual pores. Given hourglass shape of the TEM-drilled nanopores, the effective thickness is estimated as one-third of membrane thickness, which fits well with the measured ionic conductance of TEM-drilled SiN pores and was previously confirmed by TEM tomography.[13, 65] When fabricating a closely packed array, the electron beam (TEM mode) is operated at a small region. Thus, the confined region undergoes concurrent thinning by the electron beam sputtering, and the effective thickness could be even thinner.

*Conductance of single nanopore.* The resistor model of conductance for single nanopore with considering the contribution of surface charge can be described as[66]

$$G_{1,s} = \kappa_b \left( \frac{4L_\mathrm{p}}{\pi {d_\mathrm{pore}}^2} \times \frac{1}{1+\frac{4\mathrm{Du}}{d_\mathrm{pore}}} + \frac{1}{d_\mathrm{pore} + \mathrm{Du}} \right)^{-1} \tag{2}$$

where $\kappa_b$ is the bulk ionic conductivity. $d_{\mathrm{pore}}$ and $L_{\mathrm{p}}$ are the pore diameter and effective pore thickness, respectively. $\mathrm{Du} = \frac{|\sigma_s|/e}{2C_b}$ is the Dukhin length. $\sigma_s$ is the surface charge density. $e$ is the elementary charge and $C_b$ is the bulk concentration. For the dashed curve in **Figure 3**, we employ linear scaling of above equation, that is, conductance of NPA equals to $G_N$=$N$*$G_{1,\mathrm{s}}$.

*Ion Current Measurement.* Prior to ionic measurements, NPA membranes were cleaned with boiling piranha solution (1:3 v/v $H_2O_2$:$H_2SO_4$) for 20 minutes aid in pore wetting. Current-voltage measurement was performed with Keithley 2470 source meter (Keithley Tektronix, Beaverton, OR, USA) or Elements Nanopore Reader (Elements SRL, Italy). NPA membranes were sealed in a customized flow cell with Ag/AgCl electrodes in each chamber. The KCl solutions were tris-buffered at pH 8.2. When changing the concentration for measurements, the flow cell is first rinsed with DI water for at least two rounds, followed by a rinse with the target concentration, after which both chambers are filled with the solution of the target concentration.

*Theoretical Modeling.* 3D simulations are performed by solving Poisson-Nernst-Planck (PNP) equations, as follows

$$-\nabla^2\phi = \frac{1}{\varepsilon_f}\sum_{i=1}^{2} F z_i c_i \qquad (3)$$

$$\nabla \cdot \left[ -D_i \nabla c_i - D_i \frac{z_i F}{RT} c_i \nabla \phi \right] = 0 \qquad (4)$$

where $\varepsilon_f$, $F$, $R$, and $T$ are the fluid permittivity, Faraday constant, gas constant, and absolute temperature, respectively; $\phi$ is electric potential; $c_i$, $D_i$, and $z_i$ are the ionic concentration, diffusivity, ionic valence, respectively. We consider two ions in the bulk: potassium and chloride ions. The first and second terms in **Eq. 3** refer to diffusion and migration ionic flux, respectively. Consistent with the experiments, a square array configuration of cylindrical nanopores is considered. Two identical reservoirs are sufficiently larger than the nanopore array to ensure that all variables approach their bulk values at locations far from the array. A voltage is applied to one reservoir while the other is grounded. Therefore, the conductance of NPA can be obtained by dividing the current by the applied voltage under symmetric concentration conditions. With respect to the salt gradient, one reservoir is filled with a higher-concentration solution, while the other contains a lower-concentration solution. Typically, the SiN surface is negatively charged at pH 8. In our simulations, we assume a surface charge density of -20 mC/m$^2$ on the pore wall. 3D model requires considerable computational efforts. For instance, a simulation of a 25-pore array with 20 nm diameter and 50 nm thickness demands roughly 100 GB of memory. To lower computational efforts, PNP equations are solved only unless specified. Note that neglecting electroosmotic flow[26] will not influence the qualitative behavior of our system, where the surface charge density is relatively low. All simulations were carried out with COMSOL Multiphysics (version 5.6).

**Table 1. Comparison of ionic measurement and osmotic power in SiN-based NPA devices**

| Materials and Fabrication | Diameter (nm) | Thickness (nm) | Interpore distance (nm) | Number of pores | Quantity measured | ref |
|---|---|---|---|---|---|---|
| SiN<br>FIB-drilled | 200 | 50 | 400 | 3 to 49 | ***$G_N$* vs. *N* scaling** | 12 |
| SiN, $HfO_2$<br>EBL/RIE<br>TEM drilled | 2-80 | 10 to 100 | 50-100 | 1 to 150000 | Stability of $G_N$ vs. time | 29 |
| SiN<br>TEM drilled | 3 | 10, 30 | 900, 2000 | 2 | $G_N$ of two pores<br><br>DNA translocation | 14 |
| hBN/SiN<br>AFM-tip breakdown | 10 | 12 | 200-1000 | 9 | Osmotic power vs. interpore distance | 21 |
| SiN<br>EBL/RIE | 50-100 | 50 | 1000 | 2 to 1600 | Osmotic power vs. *N* scaling | 24 |
| SiN<br>TEM-drilled | ~10 | 30 | 15-200 | 2, 3 | $G_N$ vs. interpore distance, concentration<br><br>Surface modification | 15 |
| SiN<br>EBL/RIE | 60 | 30 | 1000 | 64 | Osmotic power vs. gate voltage | 63 |
| SiN<br>FIB-drilled | 30 | 30 | ~300-1500 | 1 to 49 | Osmotic power vs. porosity<br><br>Gating-voltage control | 62 |
| SiN<br>EBL/RIE<br>TEM drilled | 2 to 20 | 20 to 50 | 20 to 1000 | 4 to 10000 | ***$G_N$* vs. *N* scaling**, interpore distance, concentrations<br><br>Osmotic power vs. *N* scaling | this work |

## ASSOCIATED CONTENT

### Supporting Information.

Supporting Information is available free of charge at http://pubs.acs.org.

Characterization of NPA with cold development (Figure S1); TEM images for NPA (Figures S2 and S3); Index of NPAs fabricated by TEM-drilled method (Table S1); Variation of NPA conductance with interpore distance (Figure S4); Theoretical results: Normalized conductance (Figure S5), Electric field profile (Figure S6), current rectification for thicker membrane (Figures S8 and S9); Additional current rectification measurement (Figure S7); Pore diameter characterization (Figure S10)

### Corresponding Author

*Email: drndic@physics.upenn.edu

## NOTES

The authors declare no competing financial interest.

## ACKNOWLEDGMENTS

This work was partially funded by NSF grants 2002477, 190504, NIH grant R21 HG0101536 at the University of Pennsylvania. This work was partially performed at the Singh Center for Nanotechnology, an NNCI member supported by NSF Grant NNCI-2025608. C.-Y.L. and M.D. conceived the experiments. C.-Y.L. performed fabrication, TEM, ionic measurements, data analysis, and theoretical modeling. C.-Y.L. gratefully acknowledges Dr. Yung-Chien Chou for valuable discussions regarding array fabrication. C.-Y.L. is grateful to Dr. Douglas Yates for TEM assistance, the late Dr. Hiromichi Yamamoto for EBL assistance, and Chen Lo for graphic design. C.-Y.L. and M.D. discussed the manuscript and contributed throughout the writing process.

TOC Graphic

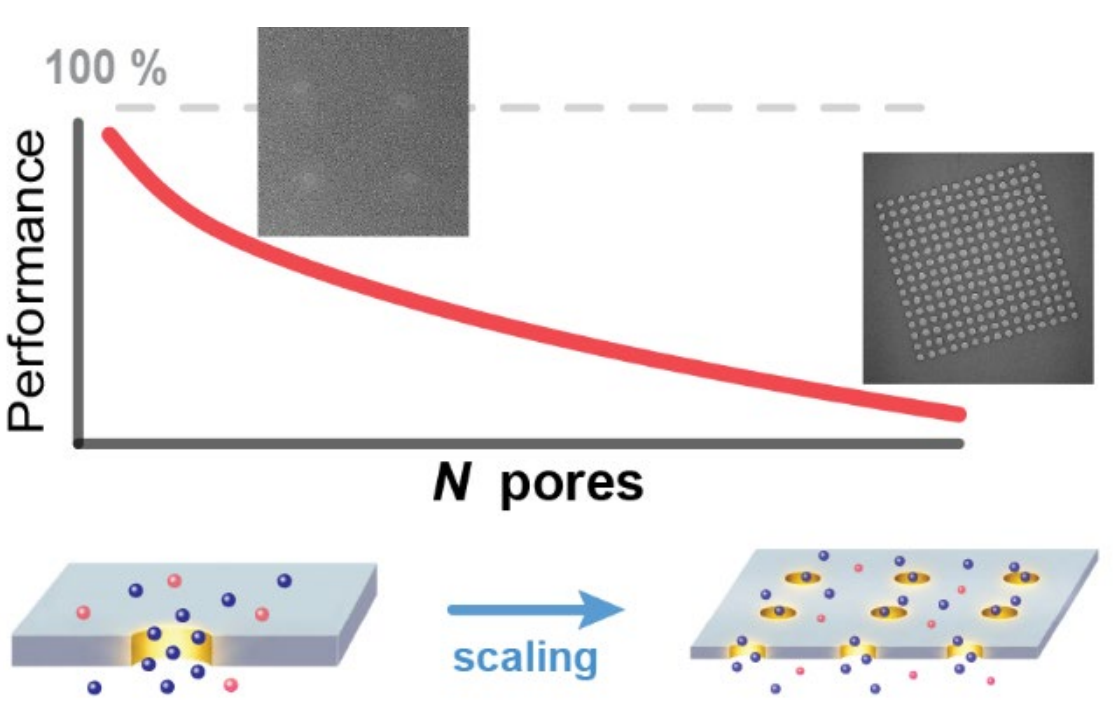